\documentclass[aps,showpacs,preprint]{revtex4}
\usepackage{graphicx}
\begin{document}
\title{Speedup of quantum state transfer by three- qubit interactions:
Implementation by nuclear magnetic resonance
 \footnote{
Jingfu Zhang: zhangjfu2000@yahoo.com, Jingfu@e3.physik.uni-dortmund.de\\
Xinhua Peng: xinhua@e3.physik.uni-dortmund.de\\
Dieter Suter: Dieter.Suter@uni-dortmund.de}}
\author{Jingfu Zhang $^{1,2}$, Xinhua Peng $^{1}$, and Dieter Suter $^{1}$}
\address{$^{1}$Fachbereich Physik, Universit$\ddot{a}$t Dortmund, 44221 Dortmund, Germany\\
$^{2}$Department of Physics, Tsinghua University, Beijing, 100084,
P R China}
\date{\today}
\begin{abstract}
Universal quantum information processing requires single-qubit
rotations and two-qubit interactions as minimal resources. A
possible step beyond this minimal scheme is the use of three-qubit
interactions. We consider such three-qubit interactions and show
how they can reduce the time required for a quantum state transfer
in an $XY$ spin chain. For the experimental implementation, we use
liquid-state nuclear magnetic resonance (NMR), where three-qubit
interactions can be implemented by sequences of radio-frequency
pulses.
\end{abstract}

\maketitle
\section{Introduction}
Quantum computers are capable of solving some computational problems
efficiently for which no efficient classical algorithms are known.
Examples include the factorization of large numbers
\cite{Shor94}, searching unsorted databases \cite{Grover97}, and
simulating quantum systems \cite{Feynman82,Somma02}.
While this advantage originates from a different scaling behavior
compared to classical computers, rather than a higher clock speed,
the time required for a single gate operation remains a critical issue:
Reliable quantum computation becomes possible only if
a sufficiently large number of gate operations can be completed
within the decoherence time of the system.

An important element of many quantum information processing
operations is the transfer of a quantum state
$\alpha|0\rangle+\beta|1\rangle$ from one qubit to another
\cite{Bose03}.
We will refer to this process as quantum state transfer (QST).
We thus discuss a system that is initially in state
$|\Psi\rangle_{i}=(\alpha|0\rangle+\beta|1\rangle)_{A}|\psi\rangle_{i}$,
where the qubit A is in state $\alpha|0\rangle+\beta|1\rangle$ and
the other qubits in state $|\psi\rangle_{i}$.
If we denote the QST
operation as $T$, the state transfer from A to B can be
represented as $T:|\Psi\rangle_{i}\rightarrow|\Psi\rangle_{f}
=|\psi\rangle_{f}(\alpha|0\rangle+e^{i\phi}\beta|1\rangle)_{B}$:
The final state corresponds to qubit B in state
$\alpha|0\rangle+e^{i\phi}\beta|1\rangle$ and the other qubits in
state $|\psi\rangle_{f}$. A quantum state transfer must thus
correctly transfer the amplitudes but not necessarily the phases
of the state \cite{Christandl}. No condition is imposed on the
state of the other qubits in the system.

  Currently there are three methods that can implement the QST. The first one is
quantum teleportation proposed by Bennett et al \cite{Bennett},
and has been experimentally realized in optical and liquid-state
nuclear magnetic resonance (NMR) systems \cite{Pan,Nielsen98}.
This method is based on quantum entanglement and requires quantum
measurements.
Classical communication is also needed if
one wants to determine the phase factor in the final state of
qubit B.
The second method is based on swap operations, where $T$ can
be represented as $T=\Pi S_{jl}$. $S_{jl}$ denotes a SWAP gate
that exchanges the states of qubits $j$ and $l$. To realize
$S_{jl}$, one needs external operations to control the qubits
other than qubits $j$ and $l$, such as switching on and off the
couplings between qubits $j$ or $l$ and the other qubits.

The third method uses a static spin-network
\cite{Christandl,Karbach}: the qubits are linearly connected by
Heisenberg interactions. Qubit A is initialized into state
$\alpha|0\rangle+\beta|1\rangle$ and the other qubits each into
state $|0\rangle$. Under the influence of a suitable static
coupling network, the system evolves such that qubit B ends up in
state $\alpha|0\rangle+e^{i\phi}\beta|1\rangle$. Unlike the second
method, the third method does not require spin couplings to be
switched on and off, so that it is one kind of quantum
computations with the 'always on' interactions \cite{Bose} that
avoids single-qubit operations. Hence it is easy to implement in
some solid-state systems \cite{DiVincenzo}. In this article we
concentrate on the third method. For example, the QST can be
implemented in a three- spin linear chain with the $XY$-
interactions
$\sigma_{x}^{1}\sigma_{x}^{2}+\sigma_{y}^{1}\sigma_{y}^{2}
    +\sigma_{x}^{2}\sigma_{x}^{3}+\sigma_{y}^{2}\sigma_{y}^{3}$.

The initial state is chosen as
$(\alpha|0\rangle+\beta|1\rangle)_{A}|00\rangle$ by setting spin
$1$ at the location A into state $\alpha|0\rangle+\beta|1\rangle$
and the other spins into state $|00\rangle$. Waiting for a period
of time $t_{0}=\frac{\pi}{2\sqrt{2}}$, one obtains the state
$|00\rangle(\alpha|0\rangle-\beta|1\rangle)_{B}$, which means that
the spin $3$ at location B is now in state
$\alpha|0\rangle-\beta|1\rangle$.
Both the initial state and the
final state are product state. However the middle state can be an
entangled states.
The relation between quantum entanglement and the
QST in the spin-network has been well discussed in Ref.
\cite{Sun}. In the three spin chain the maximum transfer distance
is 2. If one transfers a state over longer distance, one needs to
design and generate the couplings between spins in a linear chain,
or expand the chain into a spin network through introducing the
additional spins. The details can be found in Refs.
\cite{Christandl}.

Like other quantum information processing tasks, QST can be
effected with a minimum set of gates \cite{physica}, typically
consisting of single- qubit rotations and CNOT gates that can be
implemented through two- qubit interactions
\cite{Deutsch85,Somma02}. An additional possible resource are
three- spin interactions \cite{Baxter}. Effective three-particle
interactions exist in some real physical systems, for example in
optical lattices constructed of equilateral triangles
\cite{PachosPRL}. The spin Heisenberg chain with three-spin
interactions can exhibit interesting phase transition phenomena,
such as incommensurate phases \cite{Tsvelik,Frahm}, chiral phase
transitions \cite{Cruz}, or a quantum entanglement phase
transition\cite{Lou,Yang}.

The three- spin interaction that we consider corresponds to a
coupling between next- nearest- neighbors controlled by the middle
spin \cite{PachosPRL}. It is a rare resource in some quantum
systems. In this article, we use the three- spin interactions in
the spin $XY$- chain to increase the speed of the QST, and
quantitatively describe the advantages obtained by using such a
resource.

While nature does not provide three-spin interactions between nuclear spins,
they can be simulated quite readily in liquid-state NMR \cite{Tseng99}.
For this purpose, one combines the natural two-spin interactions of the type
$J_{mn}\sigma_{z}^{m}\sigma_{z}^{n}$, where $\sigma_{z}^{m}$
denotes the $z$- component of the Pauli matrix for spin $m$, and
$J_{mn}$ denotes the coupling constant between spins $m$ and $n$.
In this work, we use this approach
to generate an effective Hamiltonian with variable three-qubit
coupling strength to assess the speed-up of the QST due to
three-qubit interactions.

\section{$XY$ spin chain with three-spin interactions}\label{sect2}

\subsection {System and hamiltonian}

To test the speed-up of a state-transfer operation by three-spin
interactions, we consider a three spin $XY$ chain, which is
described by the Hamiltonian
 \begin{equation}\label{xy}
    H_{XY3}=(\sigma_{x}^{1}\sigma_{x}^{2}+\sigma_{y}^{1}\sigma_{y}^{2}
    +\sigma_{x}^{2}\sigma_{x}^{3}+\sigma_{y}^{2}\sigma_{y}^{3})+
    \frac{\lambda}{2}(\sigma_{x}^{1}\sigma_{z}^{2}\sigma_{y}^{3}-
    \sigma_{y}^{1}\sigma_{z}^{2}\sigma_{x}^{3}) .
\end{equation}
Here, $\sigma_{x/y/z}^{j}(j=1,2,3)$ are the Pauli matrices
and we have set  $\hbar$ and the coupling constant
for the two-spin terms to one.
To find an analytical expression for the time evolution of this system
and determine the conditions for state transfer, we write the Hamiltonian (1)
as a sum of two commuting parts,  $H_{XY3}= C + D$, where
\begin{equation}
C=\sigma_{x}^{1}\sigma_{x}^{2}+\sigma_{y}^{2}\sigma_{y}^{3}+
\frac{\lambda}{2}\sigma_{x}^{1}\sigma_{z}^{2}\sigma_{y}^{3},
\hspace{1.0 cm}
D=\sigma_{y}^{1}\sigma_{y}^{2}+\sigma_{x}^{2}\sigma_{x}^{3}-
\frac{\lambda}{2}\sigma_{y}^{1}\sigma_{z}^{2}\sigma_{x}^{3}.
\end{equation}

\subsection {Propagator and transfer speed}\label{2b}

This decomposition shows directly that this Hamiltonian generates a periodic time-evolution:
defining $k=\sqrt{2+\frac{\lambda^{2}}{4}}$, we find $C^2 = D^2 = k^2 I$
and therefore
\begin{equation}\label{Uxy}
   U(t)=e^{-iH_{XY3} t}
   = e^{-i t C }  e^{-i t D }
   = [\cos(kt)I-i\frac{\sin(kt)}{k}C] [\cos(kt)I-i\frac{\sin(kt)}{k}D] .
\end{equation}
For times $t = n\tau = n\pi/k $ ($n$ integer), the propagator
returns to unity, $U(\tau) = I$.

The matrix representation of the propagator is
\begin{equation}\label{Um}
    U(t)=\left (
    \begin{array}{cccccccc}
      1& 0 & 0 & 0 & 0 & 0 & 0 & 0 \\
      0 & \frac{(2kc)^{2}-(\lambda s)^{2}}{4k^{2}} & -i\frac{2kcs+\lambda s^{2}}{k^{2}} & 0 &
       \frac{k\lambda cs-2s^{2}}{k^{2}} & 0 & 0 & 0 \\
      0 & -i\frac{2kcs-\lambda s^{2}}{k^{2}} & \frac{k^{2}-4s^{2}}{k^{2}} & 0
      & -i\frac{2kcs+\lambda s^{2}}{k^{2}} & 0 & 0 & 0 \\
      0 & 0 & 0 & \frac{(2kc)^{2}-(\lambda s)^{2}}{4k^{2}} & 0 & -i\frac{2kcs-\lambda s^{2}}{k^{2}}
      & -\frac{2s^{2}+k\lambda cs}{k^{2}} & 0 \\
      0 & -\frac{2s^{2}+k\lambda cs}{k^{2}} & -i\frac{2kcs-\lambda s^{2}}{k^{2}} & 0
      & \frac{(2kc)^{2}-(\lambda s)^{2}}{4k^{2}} & 0 & 0 & 0 \\
      0 & 0 & 0 & -i\frac{2kcs+\lambda s^{2}}{k^{2}}
       & 0 & \frac{k^{2}-4s^{2}}{k^{2}} & -i\frac{2kcs-\lambda s^{2}}{k^{2}} & 0 \\
      0 & 0 & 0 & \frac{k\lambda cs-2s^{2}}{k^{2}}& 0 & -i\frac{2kcs+\lambda s^{2}}{k^{2}}
       & \frac{(2kc)^{2}-(\lambda s)^{2}}{4k^{2}} & 0 \\
      0 & 0 & 0 & 0 & 0 & 0 & 0 & 1 \\
    \end{array}\right ),
\end{equation}
where $c\equiv\cos(kt)$, and $s\equiv\sin(kt)$.

The propagator generates a state transfer at times $t_{QST} =
\arcsin\sqrt{\frac{8+\lambda^{2}}{8+2\lambda^{2}}}/k$: If $\lambda
\geq 0$, it effects a transfer from qubit 3 to qubit 1, for
negative 3-qubit coupling constant in the opposite direction. In
both cases, the periodicity of the overall evolution implies that
the reverse transfer occurs at time $t = \pi/k - t_{QST}$. The
corresponding propagators are
\begin{equation}\label{UQST13}
    U(t_{QST_{1\rightarrow 3}})=\left (
    \begin{array}{cccccccc}
      1& 0 & 0 & 0 & 0 & 0 & 0 & 0 \\
      0 & 0 & 0 & 0 &  -1 & 0 & 0 & 0 \\
      0 & i\frac{4\lambda}{\lambda^{2}+4}
      & \frac{\lambda^{2}-4}{\lambda^{2}+4} & 0 & 0 & 0 & 0 & 0 \\
      0 & 0 & 0 & 0 & 0 & i\frac{4\lambda}{\lambda^{2}+4} & \frac{\lambda^{2}-4}{\lambda^{2}+4} & 0 \\
      0 & \frac{\lambda^{2}-4}{\lambda^{2}+4} & i\frac{4\lambda}{\lambda^{2}+4} & 0 & 0 & 0 & 0 & 0 \\
      0 & 0 & 0 & 0 & 0 & \frac{\lambda^{2}-4}{\lambda^{2}+4} & i\frac{4\lambda}{\lambda^{2}+4} & 0 \\
      0 & 0 & 0 & -1 & 0 & 0 & 0 & 0 \\
      0 & 0 & 0 & 0 & 0 & 0 & 0 & 1 \\
    \end{array}\right ).
\end{equation}
for the transfer $1 \rightarrow 3$ and
\begin{equation}\label{UQST31}
    U(t_{QST_{3\rightarrow 1}})=\left (
    \begin{array}{cccccccc}
      1& 0 & 0 & 0 & 0 & 0 & 0 & 0 \\
      0 & 0 & -i\frac{4\lambda}{\lambda^{2}+4} & 0 &  \frac{\lambda^{2}-4}{\lambda^{2}+4} & 0 & 0 & 0 \\
      0 & 0 & \frac{\lambda^{2}-4}{\lambda^{2}+4} & 0 & -i\frac{4\lambda}{\lambda^{2}+4} & 0 & 0 & 0 \\
      0 & 0 & 0 & 0 & 0 & 0 & -1 & 0 \\
      0 & -1 & 0 & 0 & 0 & 0 & 0 & 0 \\
      0 & 0 & 0 & -i\frac{4\lambda}{\lambda^{2}+4} & 0 & \frac{\lambda^{2}-4}{\lambda^{2}+4} & 0 & 0 \\
      0 & 0 & 0 & \frac{\lambda^{2}-4}{\lambda^{2}+4} & 0 & -i\frac{4\lambda}{\lambda^{2}+4} & 0 & 0 \\
      0 & 0 & 0 & 0 & 0 & 0 & 0 & 1 \\
    \end{array}\right )
\end{equation}
for the transfer $3\rightarrow 1$.

\begin{figure}
\includegraphics[width=4in]{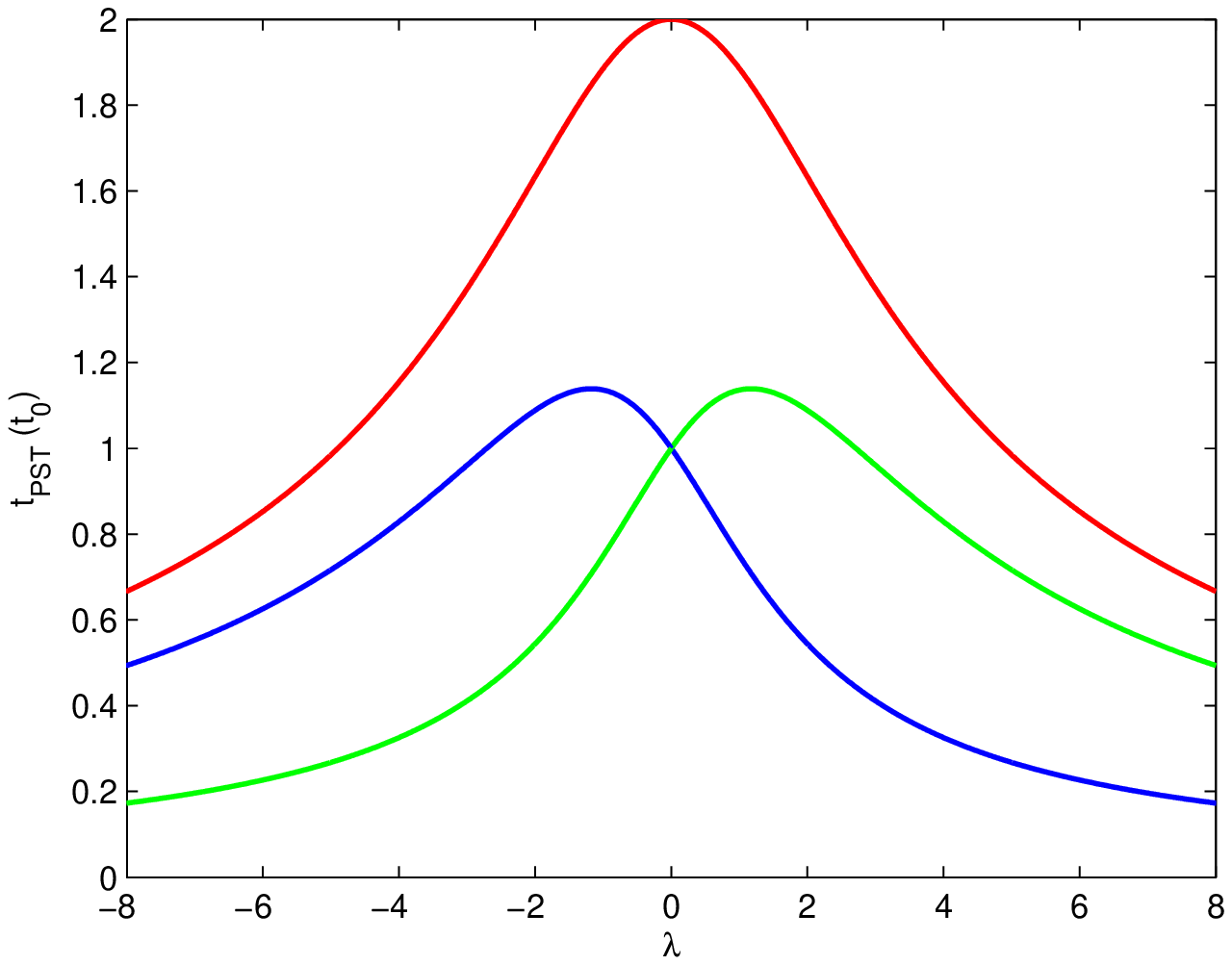}
\caption{(Color online) The duration of the QST $t_{QST_{1\rightarrow 3}}$
(green), $t_{QST_{3\rightarrow 1}}$ (blue), and
$t_{QST_{3\rightarrow 1 \rightarrow 3}}$ (red) vs $\lambda$.
The unit of the vertical axes is chosen as
$t_{0}=\frac{\pi}{2\sqrt{2}}$, normalized to the duration for
$\lambda=0$.} \label{tQST13}
\end{figure}

Figure \ref{tQST13} shows the dependence of the QST time on the
strength $\lambda$ of the three-qubit interaction. The overall
cycle time $\tau$ decreases monotonically when a three-body
coupling is added to the Hamiltonian. However, for $\lambda \neq
0$, the state transfer is no longer a simple SWAP operation, which
exchanges the states of qubits 1 and 3, but the transfer becomes
asymmetric, requiring different durations for the two directions.
While the overall cycle time $1 \rightarrow 3 \rightarrow 1$
decreases monotonically with increasing $|\lambda|$, the slower of
the two state transfers only gets faster than for $\lambda  = 0$
when $|\lambda| > 2.71199$.

\subsection{State transfer}\label{2c}

To demonstrate the state transfer, we set qubit 1 into a superposition state $\alpha|0\rangle+\beta|1\rangle$,
with the other two qubits in state $|00\rangle$.
Applying the forward state transfer Eq. (5) to this state gives
\begin{equation}
U(t_{QST_{1\rightarrow3}})(\alpha|0\rangle+\beta|1\rangle)|00\rangle
=|00\rangle(\alpha|0\rangle-\beta|1\rangle)
\end{equation}
and similar for the reverse transfer.

If we write this transfer in density operator notation, it reads
\begin{equation}\label{px1}
\left
(\begin{array}{cc}
  |\alpha|^{2} & \alpha\beta^{*} \\
  \alpha^{*}\beta & |\beta|^{2} \\
\end{array}\right)\bigotimes \left(
\begin{array}{cccc}
     1 & 0 & 0 & 0 \\
     0 &0 & 0 & 0 \\
     0 & 0 & 0 & 0 \\
     0 & 0 & 0 & 0 \\
   \end{array}
  \right)
  \rightarrow
\left(
\begin{array}{cccc}
     1 & 0 & 0 & 0 \\
     0 &0 & 0 & 0 \\
     0 & 0 & 0 & 0 \\
     0 & 0 & 0 & 0 \\
   \end{array}
  \right)
  \bigotimes \left (
 \begin{array}{cc}
  |\alpha|^{2} & -\alpha\beta^{*} \\
  -\alpha^{*}\beta & |\beta|^{2} \\
\end{array}
\right) .
\end{equation}
This result differs when the second and third qubits are initially
in different states:

\begin{eqnarray}\label{px2}
  \left
(\begin{array}{cc}
  |\alpha|^{2} & \alpha\beta^{*} \\
  \alpha^{*}\beta & |\beta|^{2} \\
\end{array}\right)\bigotimes\left(\begin{array}{cccc}
     0 & 0 & 0 & 0 \\
     0 &1 & 0 & 0 \\
     0 & 0 & 0 & 0 \\
     0 & 0 & 0 & 0 \\
   \end{array}
   \right )\rightarrow\left(\begin{array}{cccc}
     0 & 0 & 0 & 0 \\
     0 &\frac{16\lambda^{2}}{(\lambda^{2}+4)^{2}} & i\frac{4\lambda(\lambda^{2}-4)}{(\lambda^{2}+4)^{2}} & 0 \\
     0 & -i\frac{4\lambda(\lambda^{2}-4)}{(\lambda^{2}+4)^{2}} & \frac{(\lambda^{2}-4)^{2}}{(\lambda^{2}+4)^{2}} & 0 \\
     0 & 0 & 0 & 0 \\
   \end{array}
   \right )\bigotimes\left (\begin{array}{cc}
  |\alpha|^{2} & \alpha\beta^{*} \\
  \alpha^{*}\beta & |\beta|^{2} \\
\end{array}\right)
\end{eqnarray}
\begin{eqnarray}\label{px3}
  \left
(\begin{array}{cc}
  |\alpha|^{2} & \alpha\beta^{*} \\
  \alpha^{*}\beta & |\beta|^{2} \\
\end{array}\right)\bigotimes\left(\begin{array}{cccc}
     0 & 0 & 0 & 0 \\
     0 &0 & 0 & 0 \\
     0 & 0 & 1 & 0 \\
     0 & 0 & 0 & 0 \\
   \end{array}
   \right )\rightarrow\left(\begin{array}{cccc}
     0 & 0 & 0 & 0 \\
     0 &\frac{(\lambda^{2}-4)^{2}}{(\lambda^{2}+4)^{2}}& -i\frac{4\lambda(\lambda^{2}-4)}{(\lambda^{2}+4)^{2}} & 0 \\
     0 & i\frac{4\lambda(\lambda^{2}-4)}{(\lambda^{2}+4)^{2}} & \frac{16\lambda^{2}}{(\lambda^{2}+4)^{2}} & 0 \\
     0 & 0 & 0 & 0 \\
   \end{array}
   \right )\bigotimes\left (\begin{array}{cc}
  |\alpha|^{2} & \alpha\beta^{*} \\
  \alpha^{*}\beta & |\beta|^{2} \\
\end{array}\right)
\end{eqnarray}
\begin{equation}\label{px4}
   \left
(\begin{array}{cc}
  |\alpha|^{2} & \alpha\beta^{*} \\
  \alpha^{*}\beta & |\beta|^{2} \\
\end{array}\right)\bigotimes\left(\begin{array}{cccc}
     0 & 0 & 0 & 0 \\
     0 &0 & 0 & 0 \\
     0 & 0 & 0 & 0 \\
     0 & 0 & 0 & 1 \\
   \end{array}
   \right )\rightarrow\left(\begin{array}{cccc}
     0 & 0 & 0 & 0 \\
     0 &0 & 0 & 0 \\
     0 & 0 & 0 & 0 \\
     0 & 0 & 0 & 1 \\
   \end{array}
   \right)\bigotimes\left
(\begin{array}{cc}
  |\alpha|^{2} & -\alpha\beta^{*} \\
  -\alpha^{*}\beta & |\beta|^{2} \\
\end{array}\right).
\end{equation}
The phase of the superposition in the transferred state contains thus
information on the state of the other qubits.


\subsection{Mixed states and parallel implementation}\label{2d}

These different cases can be implemented in parallel by using a
mixed initial state \cite{Datta05}. By choosing
$\alpha=\beta=1/\sqrt2$ and adding the four initial states in Eqs.
(\ref{px1}-\ref{px4}), we obtain
$$ \frac{1}{2}\left
(\begin{array}{cc}
  1 & 1 \\
  1 & 1 \\
\end{array}\right)\bigotimes \left(
\begin{array}{cccc}
     1 & 0 & 0 & 0 \\
     0 &1 & 0 & 0 \\
     0 & 0 & 1 & 0 \\
     0 & 0 & 0 & 1 \\
   \end{array}
  \right)
= \frac{1}{2}(\sigma_{x}^{1}+I^{1}) \otimes I^2 \otimes I^3 .
$$
In the following, we will ignore the unit operator on qubits that
are in a superposition state.
The state transfer acting on this initial state generates then
\begin{equation}\label{mix13x}
 U(t_{QST_{1\rightarrow 3}})\sigma_{x}^{1}I^{2}I^{3}
 U(t_{QST_{1\rightarrow 3}})^{\ddag}=-\sigma_{z}^{1}
 \sigma_{z}^{2}\sigma_{x}^{3}
\end{equation}
For related initial conditions, we find
\begin{equation}\label{mix13y}
 U(t_{QST_{1\rightarrow 3}})\sigma_{y}^{1}I^{2}I^{3}U(t_{QST_{1\rightarrow
 3}})^{\ddag}=-\sigma_{z}^{1}\sigma_{z}^{2}\sigma_{y}^{3}
\end{equation}
\begin{equation}\label{mix13z}
 U(t_{QST_{1\rightarrow 3}})\sigma_{z}^{1}I^{2}I^{3}U(t_{QST_{1\rightarrow
 3}})^{\ddag}=I^1I^2\sigma_{z}^{3}.
\end{equation}
Obviously the different phases that we found in the state transfer
for the pure initial states result in the introduction of
correlations when a mixed initial state is used. Only if the
initial state is not a superposition state (Eq. \ref{mix13z}), do
we find a state transfer that does not entangle the transferred
state with the other states.

\section{Implementation in an NMR quantum computer}\label{real}

The nuclear spin system that we use to implement the stepped-up
QST has the natural Hamiltonian
\begin{equation}\label{nmr}
  H=-\pi\sum_{i=1}^3 \nu_{i}\sigma_{z}^{i}
  +\frac{\pi}{2} J_{12}\sigma_{z}^{1}\sigma_{z}^{2}
  +\frac{\pi}{2} J_{23}\sigma_{z}^{2}\sigma_{z}^{3}
\end{equation}
where $\nu_{i}$ denotes the resonance frequency of spin $i$.
Considering this system as a quantum simulator of the Heisenberg spin chain
described by the Hamiltonian (1), we generate an effective evolution (3)
by an appropriate sequence of radio frequency pulses.
While it is relatively straightforward to generate each of the terms of the Hamiltonian (1),
they do not commute with each other.
A sequential generation of the different terms therefore does not produce the correct
overall evolution.
Two different approaches allow one to generate such an evolution:
\begin{itemize}
\item Each term is implemented for a very short duration. In this
limit, the corresponding propagators are close to the unit
operator and the noncommuting terms appear only in second order
\cite{Vandersypen}.
\item The evolutions $U_{C}(t)=e^{-itC}$ and
$U_{D}(t)=e^{-itD}$ are written as a product such that each factor can be
implemented directly.
\end{itemize}
For the purpose of this paper, we have chosen the second approach.

\subsection{Decomposing $U(t)$}

\begin{figure}
\includegraphics[width=2in]{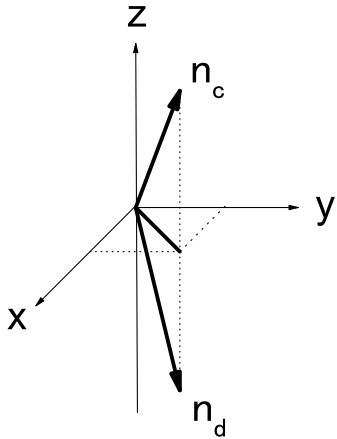}
\caption{The frame for operations $U_{C}(t)$ and $U_{D}(t)$. The
vectors ${\bf n_{c}}$ and ${\bf n_{d}}$ denote the directions of
the rotation axes for the two operations, respectively.
They
are tilted from the $z$ axis by the angles $\theta_{c}$ and
$\theta_{d}=\pi-\theta_{c}$.
The projections of ${\bf n_{c}}$ and
${\bf n_{d}}$ into the $xy$- plane are identical and indicated by a black line.
The angles between the projection and the $x$ and $y$ axes are $\pi/4$.}
\label{rotation}
\end{figure}

A suitable decomposition of $U_{C}(t)$ uses the three operators
$L_{x}^{C}\equiv
\sigma_{x}^{1}\sigma_{x}^{2}/2$,
  $L_{y}^{C}\equiv \sigma_{y}^{2}\sigma_{y}^{3}/2$, and
$L_{z}^{C}\equiv \sigma_{x}^{1}\sigma_{z}^{2}\sigma_{y}^{3}/2$.
These operators can be viewed as the three components of an
angular momentum vector $\bf{L^{C}}$, because they
satisfy the cyclic commutation relations
$[L_{x}^{C},L_{y}^{C}]=iL_{z}^{C}$ and cycl. \cite{Zhang05}.
In terms of these operators, $U_{C}(t)$ becomes
\begin{equation}\label{UAr}
    U_{C}(t)=e^{-i(\frac{2\sqrt{2}}{\sin \theta_{c}}t)(\bf{L^{C}\cdot n_{c})}}
    =e^{-i(\frac{2\sqrt{2}}{\sin \theta_{c}}t)(\frac{\sin \theta_{c}}{\sqrt{2}}L_{x}^{C}
    +\frac{\sin \theta_{c}}{\sqrt{2}}L_{y}^{C}+\cos\theta_{c}L_{z}^{C})}
\end{equation}
where $\tan\theta_{c}=\frac{2\sqrt{2}}{\lambda}$, and the vector
${\bf n_{c}}=(\frac{\sin\theta_{c}}{\sqrt{2}},\frac{\sin\theta_{c}}{\sqrt{2}},\cos\theta_{c})$
gives the direction of the rotation axis for $U_{C}(t)$, as
shown in Figure \ref{rotation}.
Using angular momentum theory, we rewrite this as
\begin{eqnarray}\label{UCf}
 U_{C}(t)&=&e^{-i \frac{\pi}{4}L_{z}^{C}}
 e^{i(\frac{\pi}{2}-\theta_{c})L_{y}^{C}}
 e^{-i(\frac{2\sqrt{2}}{\sin \theta_{c}}t)L_{x}^{C}}
 e^{-i(\frac{\pi}{2}-\theta_{c})L_{y}^{C}}
 e^{i\frac{\pi}{4}L_{z}^{C}}.
\end{eqnarray}

In a completely analogous way, we define
  $L_{x}^{D}\equiv
\sigma_{x}^{2}\sigma_{x}^{3}/2$,
  $L_{y}^{D}\equiv \sigma_{y}^{1}\sigma_{y}^{2}/2$, and
$L_{z}^{D}\equiv \sigma_{y}^{1}\sigma_{z}^{2}\sigma_{x}^{3}/2$ as
the three components of the angular momentum vector $\bf{L^{D}}$.
In terms of these operators, $U_{D}$ becomes
\begin{eqnarray}\label{UDf}
 U_{D}(t)=e^{-itD}=e^{-i \frac{\pi}{4}L_{z}^{D}}
 e^{-i(\theta_{d}-\frac{\pi}{2})L_{y}^{D}}
 e^{-i(\frac{2\sqrt{2}}{\sin \theta_{d}}t)L_{x}^{D}}
 e^{i(\theta_{d}-\frac{\pi}{2})L_{y}^{D}}
 e^{i\frac{\pi}{4}L_{z}^{D}},
\end{eqnarray}
where $\theta_{d}=\pi-\theta_{c}$.

While the two-spin terms $L_x$ and $L_y$ in  Eqs. (\ref{UCf}) and
(\ref{UDf}) are relatively easy to implement, the three-spin terms
$L_z$ are less straightforward. We re-write them as
\begin{equation}\label{xzy67}
    e^{i\eta L_{z}}=e^{i\frac{\pi}{2}L_{y}}e^{i\eta L_{x}} e^{-i\frac{\pi}{2}L_{y}} ,
    \label{e.xyz}
\end{equation}
where $\eta$ is an arbitrary  real number.
Alternatively, we may transform the propagators
$e^{i\eta L_{z}^{C}}$ and  $e^{i\eta L_{z}^{D}}$ as
\begin{equation}\label{xzy}
   e^{i\eta L_{z}^{C}}
   =   e^{i\frac{\eta}{2}\sigma_{x}^{1}\sigma_{z}^{2}\sigma_{y}^{3}}
    =e^{\mp i\frac{\pi}{4}\sigma_{y}^{1}}e^{\pm i\frac{\pi}{4}\sigma_{x}^{3}}
   e^{i\frac{\eta}{2}\sigma_{z}^{1}\sigma_{z}^{2}\sigma_{z}^{3}}
   e^{\pm i\frac{\pi}{4}\sigma_{y}^{1}}e^{\mp i\frac{\pi}{4}\sigma_{x}^{3}}
\end{equation}
and
\begin{equation}\label{yzx}
   e^{i\eta L_{z}^{D}}
   = e^{i\frac{\eta}{2}\sigma_{y}^{1}\sigma_{z}^{2}\sigma_{x}^{3}}
   =e^{\pm i\frac{\pi}{4}\sigma_{x}^{1}}e^{\mp i\frac{\pi}{4}\sigma_{y}^{3}}
   e^{i\frac{\eta}{2}\sigma_{z}^{1}\sigma_{z}^{2}\sigma_{z}^{3}}
   e^{\mp i\frac{\pi}{4}\sigma_{x}^{1}}e^{\pm i\frac{\pi}{4}\sigma_{y}^{3}} .
\end{equation}
and use the decomposition of
$\sigma_{z}^{1}\sigma_{z}^{2}\sigma_{z}^{3}$ into one- and
two-qubit operators \cite{Tseng99}
\begin{equation}\label{zzz67}
e^{i\frac{\eta}{2}\sigma_{z}^{1}\sigma_{z}^{2}\sigma_{z}^{3}}
=e^{i\frac{\pi}{4}\sigma_{x}^{2}}e^{-i\frac{\pi}{4}\sigma_{z}^{1}\sigma_{z}^{2}}
e^{i\frac{\pi}{4}\sigma_{y}^{2}}e^{i\frac{\eta}{2}\sigma_{z}^{2}\sigma_{z}^{3}}
e^{i\frac{\pi}{4}\sigma_{y}^{2}}e^{-i\frac{\pi}{4}\sigma_{z}^{1}\sigma_{z}^{2}}
e^{-i\frac{\pi}{2}\sigma_{y}^{2}}e^{-i\frac{\pi}{4}\sigma_{x}^{2}} .
\end{equation}
The expressions (\ref{xzy}-\ref{zzz67}) are identical to the
explicit forms of (\ref{e.xyz}). They use only single-qubit
operations $e^{i\phi\sigma_\alpha}$ and precessions under pairwise
couplings, $e^{i\xi\sigma_z^i\sigma_z^k}$, which are easy to
implement experimentally.

Without loss of generality, we discuss here only the case $\lambda
\geq0$. After some simplifications \cite{Zhang05,Somma02,DUPRA03},
Eqs. (\ref{UCf}) and (\ref{UDf}) can be represented as
\begin{eqnarray}\label{UCfTs}
 U_{C}(t)&=&e^{\mp i\frac{\pi}{4}\sigma_{y}^{1}}
 e^{\pm i\frac{\pi}{4}\sigma_{x}^{3}}
 e^{-i \frac{\pi}{8}\sigma_{z}^{1}\sigma_{z}^{2}\sigma_{z}^{3}}\nonumber\\
&\times&e^{\mp i\frac{\pi}{4}\sigma_{x}^{2}}
 e^{-i\frac{1}{2}[\frac{\pi}{2}-\arctan(\frac{2\sqrt{2}}{\lambda})]\sigma_{z}^{2}\sigma_{z}^{3}}
e^{\pm i\frac{\pi}{4}\sigma_{x}^{2}}\nonumber\\
&\times& e^{\mp i\frac{\pi}{4}\sigma_{y}^{2}}
e^{-it\sqrt{2+\frac{\lambda^{2}}{4}}\sigma_{z}^{1}\sigma_{z}^{2}}
e^{\pm i\frac{\pi}{4}\sigma_{y}^{2}}\nonumber\\
&\times&e^{\pm i\frac{\pi}{4}\sigma_{x}^{2}}
 e^{-i\frac{1}{2}[\frac{\pi}{2}-\arctan(\frac{2\sqrt{2}}{\lambda})]\sigma_{z}^{2}\sigma_{z}^{3}}
e^{\mp i\frac{\pi}{4}\sigma_{x}^{2}}\nonumber\\
&\times&
e^{i\frac{\pi}{8}\sigma_{z}^{1}\sigma_{z}^{2}\sigma_{z}^{3}}
e^{\pm i\frac{\pi}{4}\sigma_{y}^{1}} e^{\mp
i\frac{\pi}{4}\sigma_{x}^{3}},
\end{eqnarray}
and
\begin{eqnarray}\label{UDfTs}
 U_{D}(t)&=&e^{\mp i\frac{\pi}{4}\sigma_{x}^{1}}
 e^{\pm i\frac{\pi}{4}\sigma_{y}^{3}}
 e^{-i \frac{\pi}{8}\sigma_{z}^{1}\sigma_{z}^{2}\sigma_{z}^{3}}\nonumber\\
&\times&e^{\mp i\frac{\pi}{4}\sigma_{x}^{2}}
 e^{-i\frac{1}{2}[\frac{\pi}{2}-\arctan(\frac{2\sqrt{2}}{\lambda})]\sigma_{z}^{1}\sigma_{z}^{2}}
e^{\pm i\frac{\pi}{4}\sigma_{x}^{2}}\nonumber\\
&\times& e^{\pm i\frac{\pi}{4}\sigma_{y}^{2}}
e^{-it\sqrt{2+\frac{\lambda^{2}}{4}}\sigma_{z}^{2}\sigma_{z}^{3}}
e^{\mp i\frac{\pi}{4}\sigma_{y}^{2}}\nonumber\\
&\times&e^{\pm i\frac{\pi}{4}\sigma_{x}^{2}}
 e^{-i\frac{1}{2}[\frac{\pi}{2}-\arctan(\frac{2\sqrt{2}}{\lambda})]\sigma_{z}^{1}\sigma_{z}^{2}}
e^{\mp i\frac{\pi}{4}\sigma_{x}^{2}}\nonumber\\
&\times&
e^{i\frac{\pi}{8}\sigma_{z}^{1}\sigma_{z}^{2}\sigma_{z}^{3}}
e^{\pm i\frac{\pi}{4}\sigma_{x}^{1}} e^{\mp
i\frac{\pi}{4}\sigma_{y}^{3}},
\end{eqnarray}
respectively, and the three-spin terms are implemented according
to Eq. (\ref{zzz67}).

\subsection{System and pulse sequence}

\begin{figure}
\includegraphics[width=2.5in]{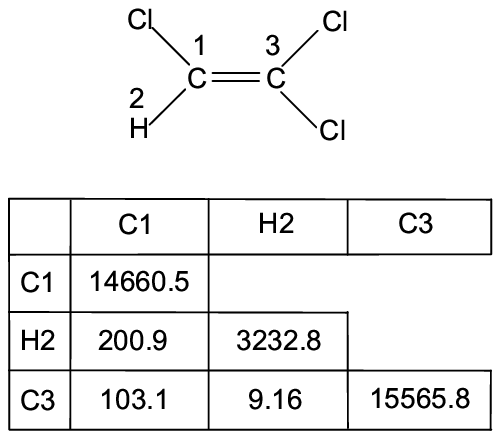}
\caption{The parameters of Carbon-13 labeled trichloroethylene
(TCE).
The diagonal terms in the table are the shifts (in Hz) of
the carbons and protons with respect to the reference frequencies
500.13MHz and 125.76MHz, respectively.
The non-diagonal terms are the coupling constants, also in Hz.}
\label{sample}
\end{figure}

\begin{figure}
\includegraphics[width=133mm]{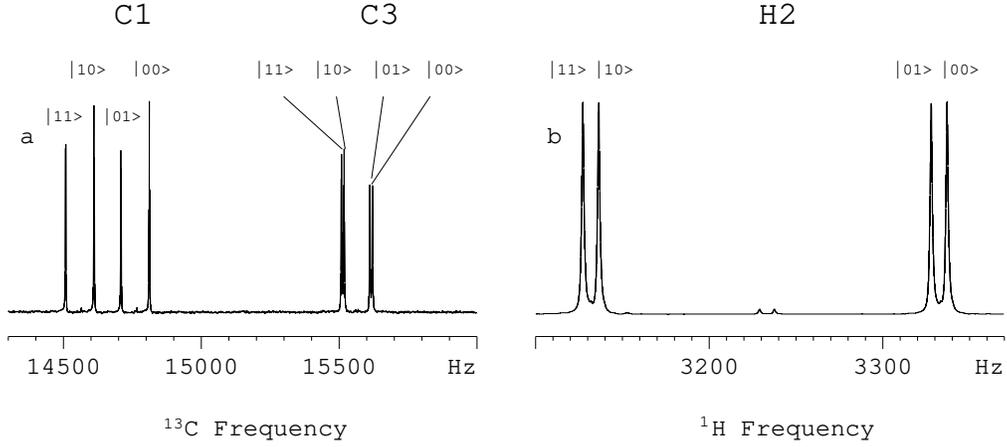}
\caption{The carbon spectrum (a) and proton spectrum (b) obtained
by applying selective readout pulses to the system in its thermal
equilibrium state.
Each qubit gives rise to four resonance lines, which correspond to
specific states of the other qubits.
The highest frequency lines always correspond to the other qubits
being in the $|00\rangle$ state, the lowest frequency lines to the
$|11\rangle$ state.
}\label{fig4}
\end{figure}

  For the experimental implementation, we used a sample of Carbon-13
labelled trichloroethylene (TCE), dissolved in d-chloroform. Data
were taken with a Bruker DRX 500 MHz spectrometer. We denote the
$^{1}$H nuclear spin as qubit 2 (H2), the $^{13}$C directly
connected to $^{1}$H is denoted as qubit 1 (C1), and the other
$^{13}$C as qubit 3 (C3). The parameters of the system and the NMR
spectra are shown in Figures \ref{sample} and \ref{fig4}. The
difference of frequency between C1 and C3 is
$\Delta\nu_{13}=905.3$Hz. The coupling constants are
$J_{13}=103.1$Hz, $J_{12}=200.9$Hz, and $J_{23}=9.16$Hz. Because
of the strongly coupled carbons \cite{Miquel} we describe the
Hamiltonian of the three-qubit system as
\begin{equation}\label{HTCE}
  H=-\pi\sum_{i=1}^3 \nu_{i}\sigma_{z}^{i}
  +\frac{\pi}{2} J_{12}\sigma_{z}^{1}\sigma_{z}^{2}
  +\frac{\pi}{2} J_{23}\sigma_{z}^{2}\sigma_{z}^{3}
  +\frac{\pi}{2} J_{13}(\sigma_{x}^{1}\sigma_{x}^{3}
  +\sigma_{y}^{1}\sigma_{y}^{3}
  +\sigma_{z}^{1}\sigma_{z}^{3}) .
\end{equation}
Since we use this system to simulate a linear chain with nearest neighbor
and three-body interactions, we do not use the coupling between qubits
1 and 3, which represent the end of the chain.

\begin{figure}
\includegraphics[width=5in]{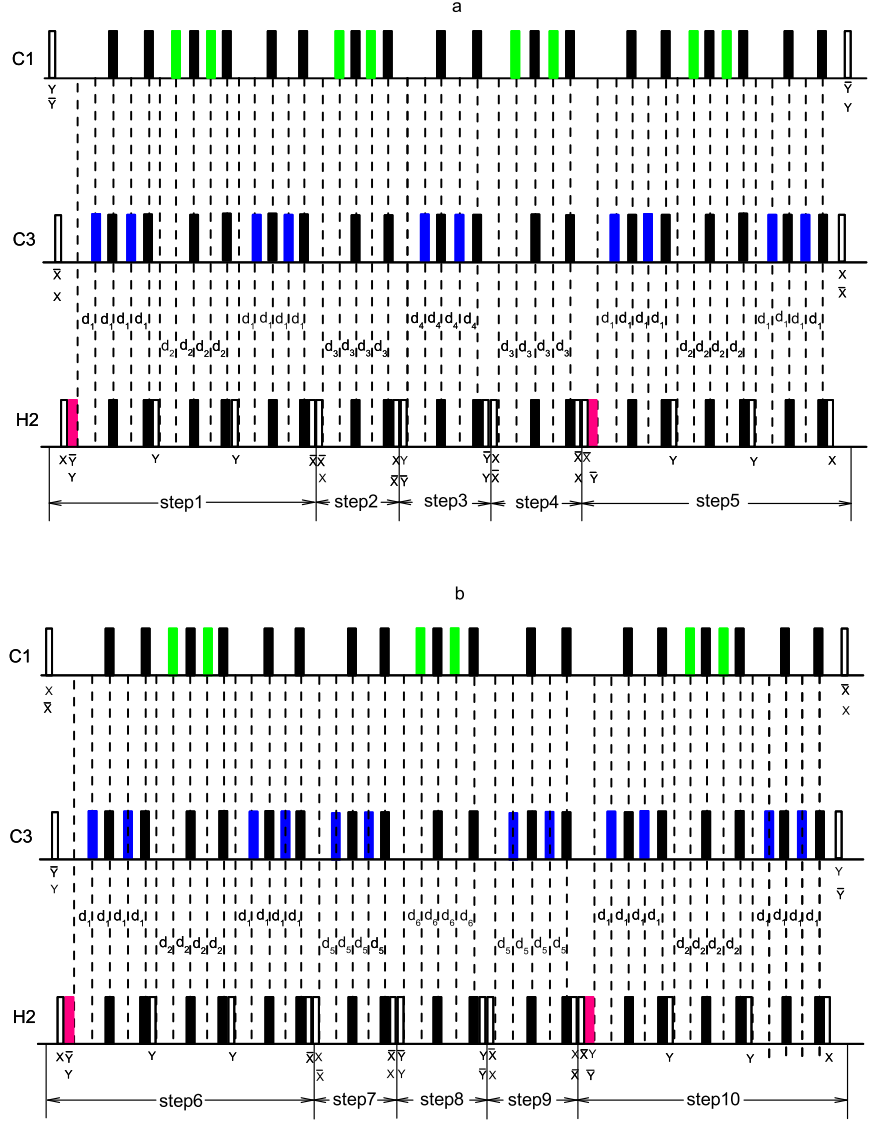}
\caption{Pulse sequences for the implementation of $U_{C}(t)$ (a)
and $U_{D}(t)$ (b). Steps $1$-$10$ correspond to the ten compound
operations separated by '$\times$' in $U_{C}(t)$ and $U_{D}(t)$,
respectively. The unfilled rectangles denote $\pi/2$ pulses, and
the filled rectangles denote $\pi$ pulses. $X$, $\overline{X}$,
$Y$ and $\overline{Y}$ below the pulses denote the $x$, $-x$, $y$,
and $-y$ directions along which the pulses are applied. Those
$\pi$ pulses for which directions are not denoted are refocusing
pulses. They are applied in pairs in which the two pulses take
opposite directions to reduce experimental errors. The durations
of the pulses applied to H2 and the non-selective pulses applied
to C1 and C3 are so short that they can be ignored. The selective
$\pi$ pulses for C1 and C3, denoted by the green or blue
rectangles, are implemented as RE-BURP \cite{REBURP} and Gauss
shaped pulses with $6.2649ms$ and $2.8252ms$ durations,
respectively. The delays are $d_{1}=\frac{9}{8J_{12}}$,
$d_{2}=\frac{1}{16J_{23}}$, $d_{3}=\frac{1}{4\pi
J_{23}}[\frac{\pi}{2}-\arctan(\frac{2\sqrt{2}}{\lambda})]$,
$d_{4}=\frac{1}{2\pi
J_{12}}(t\sqrt{2+\frac{\lambda^{2}}{4}}+2\pi)$,
$d_{5}=\frac{1}{2\pi
J_{12}}\{\pi-\frac{1}{2}[\frac{\pi}{2}-\arctan(\frac{2\sqrt{2}}{\lambda})]\}$,
$d_{6}=\frac{t}{2\pi J_{23}}\sqrt{2+\frac{\lambda^{2}}{4}}$. For
the case of $\lambda=0$, steps 2, 4, 7, and 9 are omitted.}
\label{seq}
\end{figure}

Because our quantum register contains only one proton spin, we can
implement the rotations $e^{\pm i\frac{\pi}{4}\sigma_{x/y}^{2}}$
by hard $\pi/2$ proton pulses, which are selective for qubit H2.
We denote rotations along the  $\pm x$ or $\pm y$ axis as
$[\pm\frac{\pi}{2}]_{x/y}^{2}$. The widths of such pulses are so
short that they can be considered as ideal rotations. Figures
\ref{seq} show the actual pulse sequences that we used to
implement $U_{C}$ and $U_{D}$.

Implementing spin-selective operations on the carbon spins turned
out to be difficult. We minimized experimental errors by replacing
selective pulses with non-selective pulses and free precession
periods \cite{Ryan}, using, e.g.,
\begin{equation}\label{rx1or3}
    e^{\pm i\frac{\pi}{4}\sigma_{x}^{m}}=e^{\mp i\frac{\pi}{4}\sigma_{y}^{1,3}}
    e^{i\frac{\pi}{4}\sigma_{z}^{m}}e^{\pm i\frac{\pi}{4}\sigma_{y}^{1,3}},
\end{equation}
and
\begin{equation}\label{ry1or3}
    e^{\pm i\frac{\pi}{4}\sigma_{y}^{m}}=e^{\pm i\frac{\pi}{4}\sigma_{x}^{1,3}}
    e^{i\frac{\pi}{4}\sigma_{z}^{m}}e^{\mp i\frac{\pi}{4}\sigma_{x}^{1,3}},
\end{equation}
with $m=1$ or $3$.
The $\pi/2$ rotations $e^{\pm i\frac{\pi}{4}\sigma_{x/y}^{1,3}}$
act on both carbon spins C1 and C3 and were realized by hard $\pi/2$ pulses.
The z-rotations $e^{i\frac{\pi}{4}\sigma_{z}^{m}}$ of individual qubits were
implemented by the "chemical shift rotation" method of Linden et al. \cite{Linden}.

The $\pm$ signs in Eqs. (\ref{UCfTs}) and (\ref{UDfTs}) refer to
two formally different expressions that represent the same overall
transformation. Implementing both forms and summing over the
result turned out to be very useful for suppressing experimental
artifacts arising from nonideal gate operations. When the
operations $U_C$ and $U_D$ are concatenated, it is possible to
combine the last operation of $U_C$ with the first of $U_D$ and
realize them as a hard pulse $[-\frac{\pi}{2}]_{x}^{1,3}$.

\subsection{Experimental transfer of $(|0\rangle+|1\rangle)/\sqrt{2}$}

As discussed in section \ref{2b} and shown in Figure \ref{tQST13},
the transfer from qubit 3 to 1 is always speeded up by the
three-spin interaction for $\lambda >0$. We therefore start with
this transfer, initializing the system to the state
$(|000\rangle+|001\rangle)/\sqrt{2}$.
To calculate its time evolution, we note that,
according to Eq. (\ref{Um}), the state $|000\rangle$
is an eigenstate of the Hamiltonian, $U(t)|000\rangle=|000\rangle$.
Also from Eq. (\ref{Um}), we find
 \begin{equation}\label{purex}
 U(t)|001\rangle=
 \frac{(2kc)^{2}-(\lambda s)^{2}}{4k^{2}}|001\rangle
 -i\frac{2kcs-\lambda s^{2}}{k^{2}}|010\rangle
 -\frac{2s^{2}+k\lambda cs}{k^{2}}
 |100\rangle.
\end{equation}
We monitor the progress of the state transfer by the amplitudes of
the states $|001\rangle$ and $|100\rangle$: In a superposition
with state $|000\rangle$, they correspond to $x$- magnetization of
the qubits C$3$ and C$1$, respectively.

As discussed in sections \ref{2c} and \ref{2d}, we can observe the
transfer from the 4 initial states
$|00\rangle(|0\rangle+|1\rangle)/\sqrt{2}$,
$|01\rangle(|0\rangle+|1\rangle)/\sqrt{2}$,
$|10\rangle(|0\rangle+|1\rangle)/\sqrt{2}$ and
$|11\rangle(|0\rangle+|1\rangle)/\sqrt{2}$ to their respective
final states in parallel by preparing their sum as a mixed state
$I^{1}I^{2}\sigma_{x}^{3}$ using pulse sequence
$$[\frac{\pi}{2}]_{y}^{2}-[grad]_{z}-[\frac{\pi}{2}]_{x}^{1}
-[grad]_{z}-[\frac{\pi}{2}]_{y}^{3}$$ where $[grad]_{z}$ denotes a
gradient pulse along $z$- axis. As usual \cite{deviation}, we
describe these mixed states in an operator notation that refers
only to the traceless part of the density operator. Since $[D,
\sigma_x^3] = [C, D] = 0$, the evolution of this initial condition
is determined by $C$ alone,
 \begin{eqnarray}\label{px}
 \rho_{1}(t)&=&U(t)\sigma_{x}^{3} U(t)^{\ddag}
 =U_{C}(t)\sigma_{x}^{3}U_{C}(t)^{\ddag}\nonumber\\
 &=&\frac{(2kc)^{2}-(\lambda s)^{2}}{4k^{2}}\sigma_{x}^{3}
 -\frac{2kcs-\lambda s^{2}}{k^{2}}\sigma_{y}^{2}\sigma_{z}^{3}
 -\frac{2s^{2}+k\lambda cs}{k^{2}}
 \sigma_{x}^{1}\sigma_{z}^{2}\sigma_{z}^{3} .
\end{eqnarray}

\begin{figure}
\includegraphics[width=7.5in]{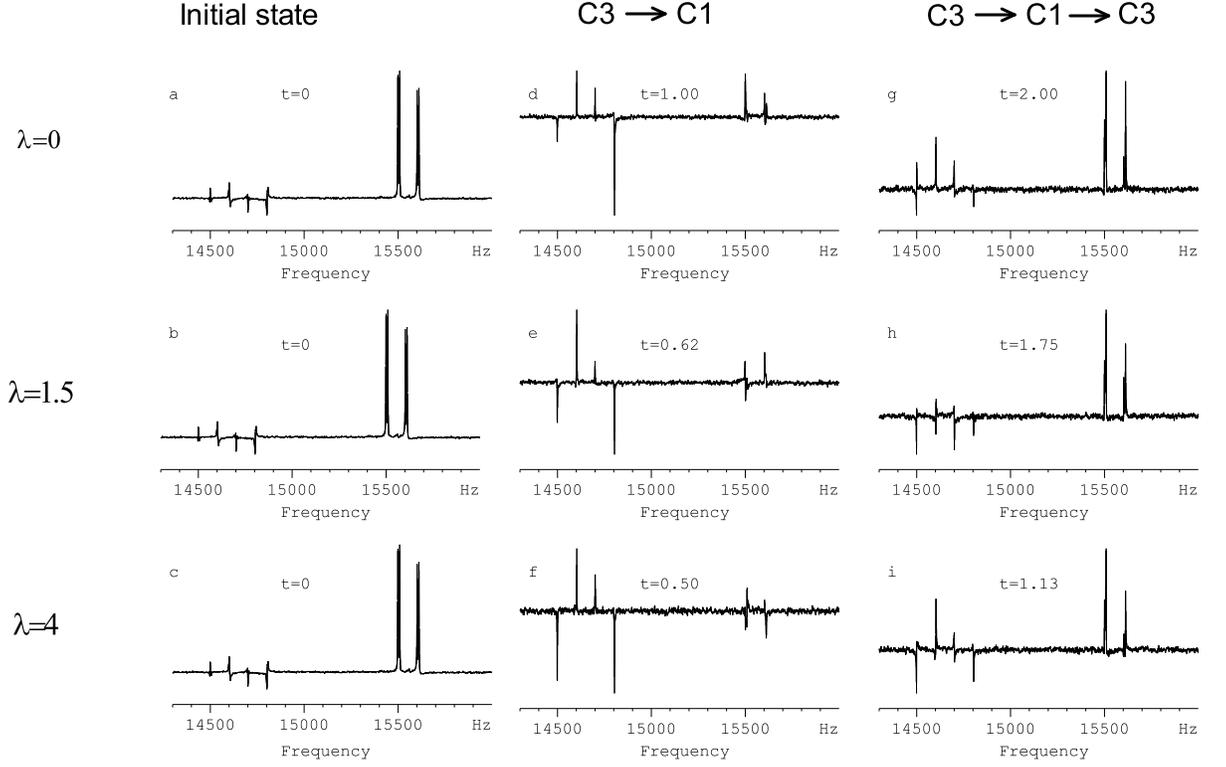}
\caption{ The experimental results demonstrating the QST.
The initial state is $\sigma _{x}^{3}$; the corresponding spectrum is
shown in the left hand column.
The results of the QST C3$ \rightarrow$ C1 are shown as Figures (d-f),
and the results of the cyclic transfer C3 $\rightarrow$ C1 $\rightarrow$ C3
are shown as Figures (g-i).
The three rows correspond to increasing three-qubit coupling strength,
$\lambda=0$, $1.5$, and $4$.
The time required for each transfer is shown in the Figures.
At $t=t_{QST_{3\rightarrow 1}}$, the three-spin system is in state
$-\sigma_{x}^{1}\sigma_{z}^{2}\sigma_{z}^{3}$,
and at $t=t_{QST_{3\rightarrow 1 \rightarrow 3}}$, the
system is in state $\sigma_{x}^{3}$.}
\label{expres}
\end{figure}

The first and last term in Eqn.(\ref{px}) correspond to directly
observable magnetization. We can therefore monitor the progress of
the quantum state transfer by simply recording the free induction
decay (FID) signal and calculating its Fourier transform.
Figures \ref{expres} show the corresponding $^{13}$C NMR spectra
observed before and after the QST, using TCE.  The initial
condition shows that the signal is concentrated on qubit C3 shown
as Figures \ref{expres} (a-c). After the transfer C3 $\rightarrow$
C1, the system is in state
$-\sigma_{x}^{1}\sigma_{z}^{2}\sigma_{z}^{3}$. The main signal is
on qubit C1, shown as Figures \ref{expres} (d-f) corresponding to
$\lambda=0$, $1.5$, and $4$, respectively. The different resonance
lines indicate that the magnetization on qubit C1 is aligned along
the positive or negative $x$- axis, depending on the state of
qubits H2 and C3. This agrees well with the prediction of Eqs.
(\ref{px1}-\ref{px4}). After the transfer C3 $\rightarrow$ C1
$\rightarrow$ C3, the system is in state $\sigma_{x}^{3}$. The
main signal returns to qubit C3, shown as Figures \ref{expres}
(g-i) corresponding to $\lambda=0$, $1.5$, and $4$, respectively.
 The time of QST C3 $\rightarrow$
C1 is measured to be $t=1.00$, $t=0.62$, and $t=0.50$, and the
time of QST C3 $\rightarrow$ C1 $\rightarrow$ C3 is measured to be
$t=2.00$, $t=1.75$, and $t=1.13$, when $\lambda=0$, $1.5$, and
$4$, respectively. Here we also use $t_{0}$ as the time unit.
Compared with the case of $\lambda=0$, the speed of QST is
increased by the three- spin interactions.
The experimental errors mainly result from the strong coupling
between the two carbons and the effects of decoherence. Moreover
the imperfection of the pulses, especially the $\pi$ pulses for
refocusing is another error source.

Similarly the process of transferring
$(|0\rangle+i|1\rangle)/\sqrt{2}$ can be observed by choosing the
initial state as $I^{1}I^{2}\sigma_{y}^{3}$ where
$\sigma_{y}=(|0\rangle+i|1\rangle)(\langle0|-i\langle1|)-I$. Using
$[C, \sigma_y^3] = 0$ we then have
\begin{eqnarray}\label{py}
 \rho_{2}(t)&=&U(t)\sigma_{y}^{3}U(t)^{\ddag}
 = U_{D}(t)\sigma_{y}^{3}U_{D}(t)^{\ddag}\nonumber\\
 &=&\frac{(2kc)^{2}-(\lambda s)^{2}}{4k^{2}}\sigma_{y}^{3}
 +\frac{2kcs-\lambda s^{2}}{k^{2}}\sigma_{x}^{2}\sigma_{z}^{3}
 -\frac{2 s^{2}+k\lambda cs}{k^{2}}
 \sigma_{y}^{1}\sigma_{z}^{2}\sigma_{z}^{3}.
\end{eqnarray}

For these initial conditions, it is thus sufficient to consider
only part of the evolution operator, generating either $U_C(t)$ or
$U_D(t)$.

\subsection{General initial conditions }

For other initial conditions, the full evolution
operator $U(t)$ is required.
As an example, we choose
$I^{1}I^{2}\sigma_{z}^{3}$ as the initial state, and obtain
\begin{eqnarray}\label{ptz}
    \rho_{3}(t)&=&U(t)\sigma_{z}^{3}U^{\dag}(t) \nonumber\\
    &=&\frac{(2kc)^{2}-(\lambda s)^{2}}{4k^{2}}[\frac{(2kc)^{2}-(\lambda s)^{2}}{4k^{2}}\sigma_{z}^{3}
        -\frac{2kcs-\lambda s^{2}}{k^{2}}\sigma_{x}^{2}\sigma_{y}^{3}
        +\frac{2s^{2}+k\lambda cs}{k^{2}}\sigma_{y}^{1}\sigma_{z}^{2}\sigma_{y}^{3}] \nonumber\\
    &+&\frac{2kcs-\lambda s^{2}}{k^{2}}[\frac{(2kc)^{2}-(\lambda s)^{2}}{4k^{2}}\sigma_{y}^{2}\sigma_{x}^{3}
      +\frac{2kcs-\lambda s^{2}}{k^{2}}\sigma_{z}^{2}
      +\frac{2s^{2}+k\lambda cs}{k^{2}}\sigma_{y}^{1}\sigma_{x}^{2}]\nonumber\\
 &+&\frac{2s^{2}+k\lambda cs}{k^{2}}[\frac{(2kc)^{2}-(\lambda s)^{2}}{4k^{2}}\sigma_{x}^{1}\sigma_{z}^{2}\sigma_{x}^{3}
      -\frac{2kcs-\lambda s^{2}}{k^{2}}\sigma_{x}^{1}\sigma_{y}^{2}+\frac{2s^{2}+k\lambda
      cs}{k^{2}}\sigma_{z}^{1}].
\end{eqnarray}

  Noting that $J_{23}$ is much smaller than $J_{12}$, one finds that
$U_{D}$ requires a longer time to complete than $U_{C}$.
For
example, when $\lambda=1.5$, $U_{C}$ requires about 340 ms for QST
from C3 to C1,  while $U_{D}$ requires about 420 ms. The effective
$T_{2}$ ($T_{2}^{*}$) of the current sample is measured to be
$0.35s$, $0.26s$, and $0.23s$ for C1, H2, and C3, respectively.
When $U_{D}$ or the full $U$ is applied, decoherence results in a
significant degradation of the experimental data.
We therefore show here only the results of the simulation.
For this purpose, we also neglect the small strong-coupling effects
between qubits C1 and C3.
The initial states are chosen as $I^{1}I^{2}\sigma_{x}^{3}$,
$I^{1}I^{2}\sigma_{y}^{3}$ and $I^{1}I^{2}\sigma_{z}^{3}$ as
respectively. Because the relevant terms in Eq. (\ref{ptz}),
are not directly observable, we apply readout pulses
$[\frac{\pi}{2}]_{y}^{1}$ and $[\frac{\pi}{2}]_{y}^{3}$
to $\rho_{3}(t)$ to obtain the observable signals of C1
and C3, respectively.

\begin{figure}
\includegraphics[width=6in]{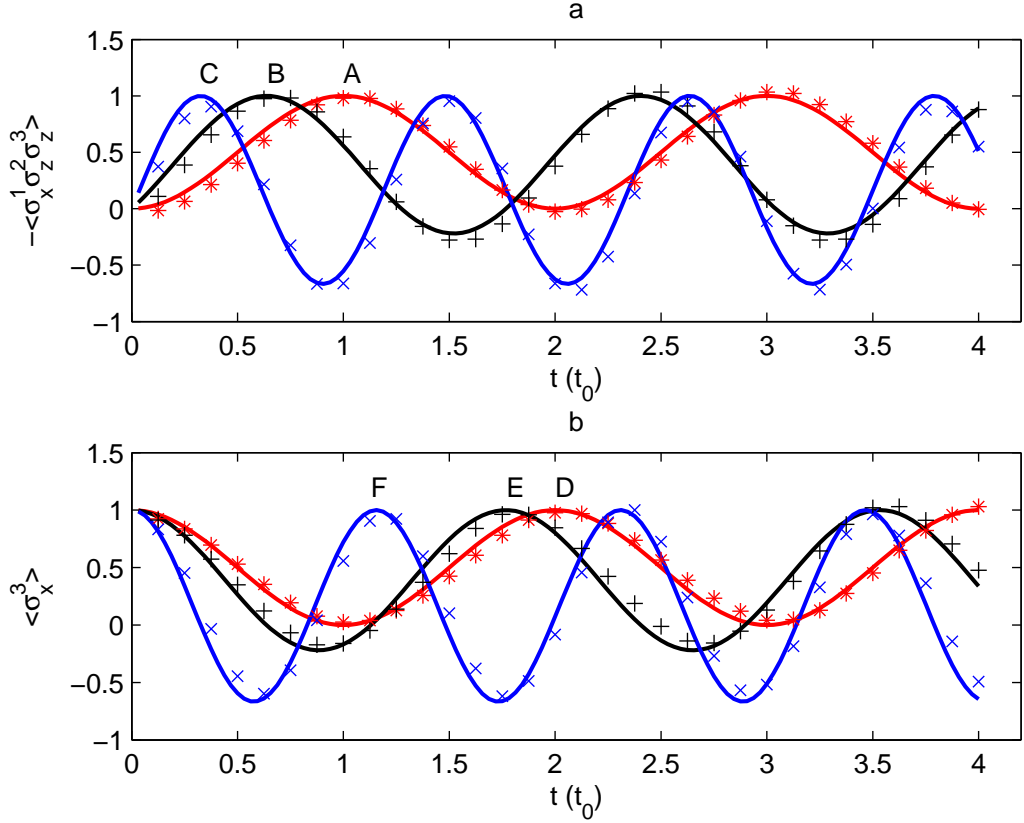}      
\caption{(Color online) Progress of the QST, starting from $\sigma
_{x}^{3}$, for different strengths of the three-body coupling. The
upper part of the figure shows the overlap of the density operator
with the target state $\sigma_x^1\sigma_z^2\sigma_z^3$ as a
function of time. The unit $t_0$ of the time axis corresponds to
the transfer time in the absence of the three-body interaction.
The data for $\lambda=0$, $1.5$, and $4$ are marked by "*", "+",
and "$\times$", respectively. The solid lines represent the
theoretical results, the individual points correspond to the
simulated data by setting TCE as the weak coupling system without
decoherence. Points A, B, and C indicate the maxima corresponding
to the transfer times C3$ \rightarrow$ C1 and the points D, E, and
F to the transfer times C3 $\rightarrow$ C1 $\rightarrow$ C3. This
clearly demonstrates the speedup of the transfer by the three-body
interaction. } \label{QSTx}
\end{figure}

\begin{figure}
\includegraphics[width=6in]{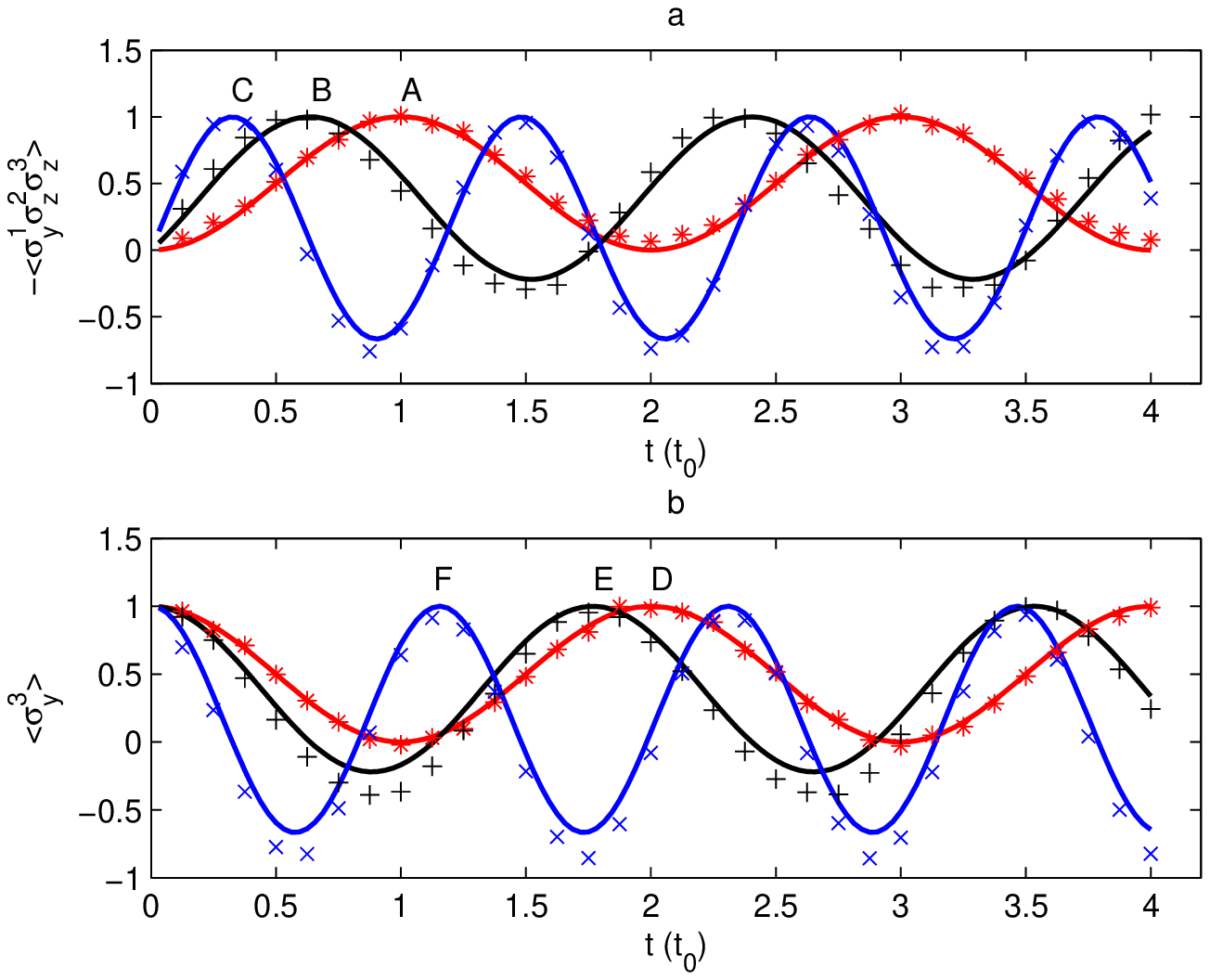}      
\caption{(Color online) Progress of the QST, starting from $\sigma
_{y}^{3}$, for different strengths of the three-body coupling. The
data
for $\lambda=0$, $1.5$, and $4$ are marked by "*", "+", and
"$\times$", respectively.
The graphs are the theoretical results, used to fit the
corresponding
data. Points A, B, and C indicate the maxima corresponding to the
transfer times C$3$ $\rightarrow$ C$1$ and points D, E, and F to
 the transfer times C$3$ $\rightarrow$ C$1$
$\rightarrow$ C$3$.}\label{QSTy}
\end{figure}

\begin{figure}
\includegraphics[width=6in]{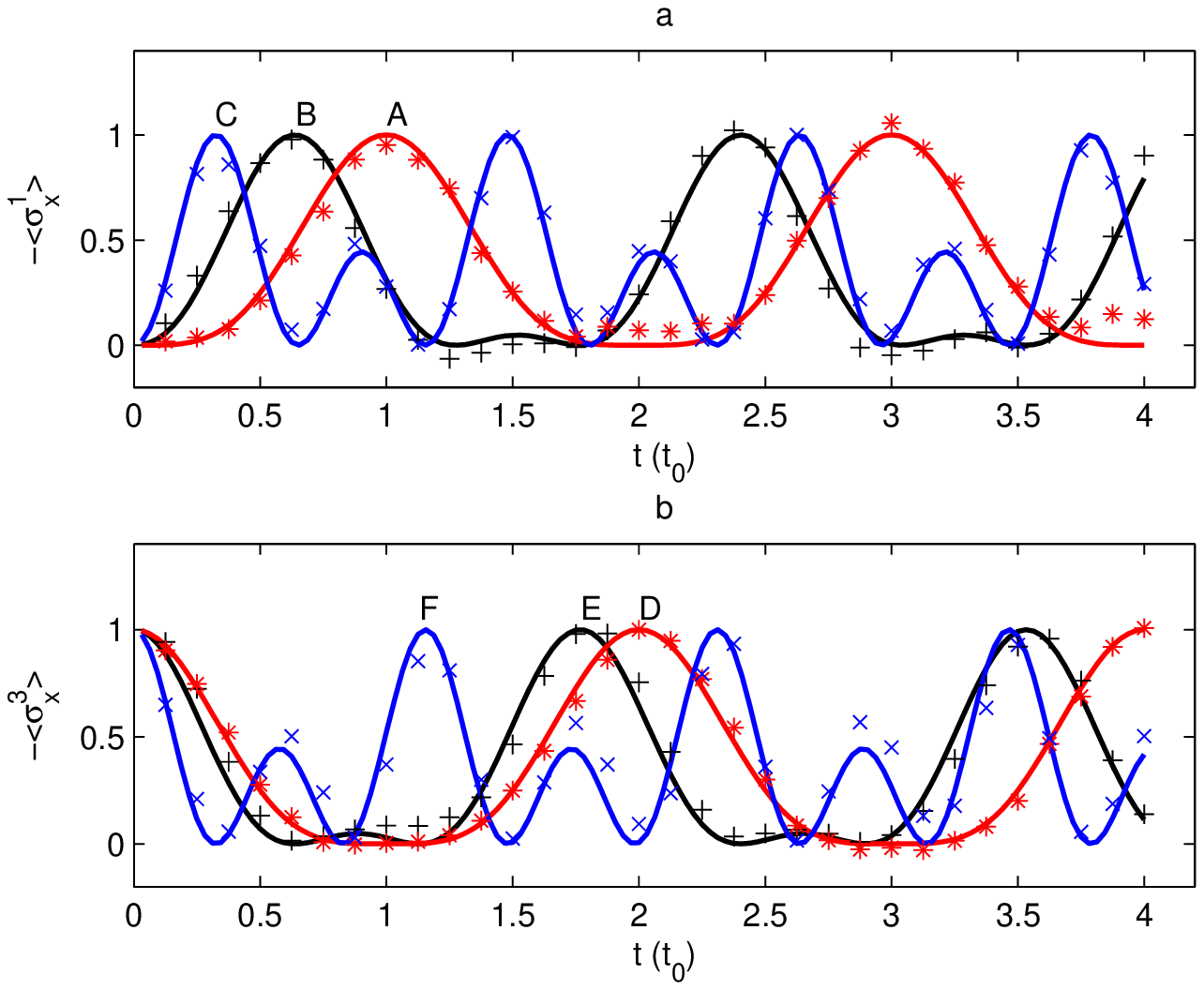}
\caption{(Color online) Progress of the QST, starting from $\sigma
_{z}^{3}$, for different strengths of the three-body coupling. The
data
for $\lambda=0$, $1.5$, and $4$ are marked by "*", "+", and
"$\times$", respectively.
The graphs are the theoretical results, used to fit the
corresponding
data. Points A, B, and C indicate the maxima corresponding to the
transfer times C$3$ $\rightarrow$ C$1$ and points D, E, and F to
the transfer times C$3$ $\rightarrow$ C$1$ $\rightarrow$
C$3$.}\label{QSTz}
\end{figure}

Figures  \ref{QSTx}-\ref{QSTz} show the progress of the QST.
For each initial state, the results for
 $\lambda=0$, $1.5$, and $4$ are given. The data points can be
 well fitted by the corresponding theoretical graphs. Points A, B, and C denote
  the maxima corresponding to the
time of QST  C$3\rightarrow$ C$1$; points D, E, and F denote the
maxima corresponding to the time of  QST C$3\rightarrow$ C$1$
$\rightarrow$ C$3$. Obviously the time required for the QST
decreases with the increase of $\lambda$.

\section{Discussion}
  The QST can also be implemented by a series of SWAP operations.
For the three- spin chain, the state of spin 1 can be transferred
to spin 3 through
\begin{equation}\label{s13}
    S_{13}=S_{12}S_{23}S_{12}=
\left (\begin{array}{cccccccc}
  1 & 0 & 0 & 0 & 0 & 0 & 0 & 0 \\
  0 & 0 & 0 & 0 & 1 & 0 & 0 & 0 \\
  0  & 0  & 1  &  0 & 0  & 0  &  0 & 0  \\
  0  &  0 &  0 & 0  &0   &   0&  1 &0   \\
  0  & 1  &  0 & 0  & 0  &  0 & 0  & 0  \\
  0  & 0  & 0  &0   &  0 & 1  &  0  &  0 \\
  0  & 0  & 0  &  1 & 0  &  0 & 0  & 0  \\
  0  & 0  &  0 & 0  & 0  &  0 & 0  & 1  \\
\end{array}\right ).
\end{equation}
When $\lambda=0$ Eqs.(\ref{UQST31}) and (\ref{UQST13}) are
equivalent to $S_{13}$ (up to some phase factors), just as
discussed in Ref. \cite{Zhang05}. When $\lambda\neq0$, however,
neither Eq.(\ref{UQST31}) nor (\ref{UQST13}) is equivalent to
$S_{13}$. The difference between the stepped-up QST and the SWAP
operation comes from the three- spin interaction, which breaks the
symmetry for exchanging spins $1$ and $3$. One can prove that when
spin $1$ and spin $3$ are exchanged, the three- spin terms in Eq.
(\ref{xy}) are changed from
$\frac{\lambda}{2}(\sigma_{x}^{1}\sigma_{z}^{2}\sigma_{y}^{3}-
    \sigma_{y}^{1}\sigma_{z}^{2}\sigma_{x}^{3})$ to
    $-\frac{\lambda}{2}(\sigma_{x}^{1}\sigma_{z}^{2}\sigma_{y}^{3}-
    \sigma_{y}^{1}\sigma_{z}^{2}\sigma_{x}^{3})$. Such asymmetry
can also explain why $t_{QST_{3\rightarrow 1}}$ differs from
$t_{QST_{1\rightarrow 3}}$.

While we have considered here only single-qubit states, it is also possible
to transfer multi-qubit states through the Heisenberg $XY$ spin
  chain \cite{Christandl,Subrahmanyam}, even entangled ones.
Such transfers can
  also be speeded up by three-spin interactions.
  For example,
  the four Bell-states $(|00\rangle\pm|11\rangle)/\sqrt{2}$,
$(|01\rangle\pm|10\rangle)/\sqrt{2}$ can be transferred from spins
2 and 3 to spins 1 and 2 by
\begin{equation}\label{Bell12}
    U(t_{QST_{1\rightarrow 3}})(|0\rangle_{1}|00\rangle\pm|11\rangle)_{23}/\sqrt{2}
    =(|00\rangle\mp|11\rangle)_{12}|0\rangle_{3}/\sqrt{2}
\end{equation}
\begin{equation}\label{Bell34}
    U(t_{QST_{1\rightarrow 3}})(|0\rangle_{1}|01\rangle\pm|10\rangle)_{23}/\sqrt{2}
    =(|01\rangle\pm|10\rangle)_{12}|0\rangle_{3}
    (\frac{\lambda^{2}-4}{\lambda^{2}+4}+i\frac{4\lambda}{\lambda^{2}+4})/\sqrt{2}.
\end{equation}
Using the analysis of section \ref{sect2}, one finds that when
$\lambda<0$, the speed of transferring the entangled states is
increased by the three-spin interactions.

\section{conclusion}
We simulated a spin XY chain with three-spin interactions,
using a three qubit NMR system.
Compared to the case where the system contains only two-spin interactions,
the three-spin interaction increases the speed of the operation.
Our results [Eqs. (\ref{UQST31}) and (\ref{UQST13})]
show that when the three- spin interactions exist, the QST is not
equivalent to the SWAP operation any more. Unlike the SWAP
operation, not all rows in the unitary evolution to realize the
QST have only one nonzero terms.


  The simulation of the XY chain with three- spin interactions
offers a possible laboratory to study the problems related to
three- spin interactions. Our techniques can simulate the chain
with arbitrary $\lambda$. In fact $\lambda$ represents the ratio
of the three-body and two- body coupling constants, because we
have set the two- body coupling constants to $1$. In a practical
sense $\lambda$ can be enhanced through increasing the three-body
couplings or decreasing the two-body couplings to speed up the
QST.
Although our results are obtained using three spin system, they are helpful
for the case of more than three spins.

\section{Acknowledgment}
  We thank Prof. Guilu Long, Prof. Jiangfeng Du and Mr. Bo Chong for helpful discussions. The
experiments were performed at the Interdisciplinary Center for
Magnetic Resonance. This work is supported by the Alexander von
Humboldt Foundation, the National Natural Science Foundation of
China under grant No. 10374010, and the DFG.



\end{document}